\documentclass[aps, prl, reprint]{revtex4-2}

\usepackage{graphicx}
\usepackage{orcidlink}
\usepackage{amsmath}
\usepackage{amssymb}
\usepackage{hyperref}
\usepackage{natbib}

\begin{document}

\title{A Self-Organized Tower of Babel: Diversification through Competition}
\author{Riz Fernando Noronha \orcidlink{0009-0007-2923-3835}}
\email[Contact author: ]{noronha@nbi.ku.dk}
\author{Kunihiko Kaneko \orcidlink{0000-0001-6400-8587}}
\affiliation{Niels Bohr Institute, University of Copenhagen, Copenhagen 2200, Denmark}

\begin{abstract}
    We introduce a minimal evolutionary model to show how local cooperation and global competition can create a transition to the diversity of communities such as linguistic groups. By using a lattice model with high-dimensional state agents and evolution under a fitness that depends on an agent's local neighborhood and global dissimilarity, clusters of diverse communities with different fitness are organized by equalizing the finesses on the boundaries, where their numbers and sizes are robust to parameters. We observe successive transitions over quasi-stationary states, as triggered by the emergence of new communities on the boundaries. Our abstract framework provides a simple mechanism for the  diversification of culture.
\end{abstract}

\maketitle


Interactions between individuals, humans, or other living organisms often have two antagonistic aspects: in one aspect, they share common information to cooperate with each other, and in the other, they keep information confidential within the group to prevent it from being exploited by other groups. The former leads to unity of individuals into a common group, whereas the latter leads to a trend to division into subgroups. Such antagonistic interactions underlie how groups with common knowledge, culture, or institution emerge,  expand, and then divide into some subgroups that share common and confidential information. 

Such antagonistic features exist even in ``language”. It is generally taken for granted that ``language has evolved as a means of communication,", which sounds intuitive and true, as the primary purpose of language is known to be a means of transferring information \cite{tomasello2010origins, pinker1990natural}. However, if only this aspect of language is taken into account, it is difficult to understand why there are so many languages that cannot communicate with each other. In terms of communication optimization, the language diversity should decrease and homogenize. Furthermore, languages have several complexities that naively would seem to hamper information transfer. At this point, there also exists a ``discommunication” aspect in language, so that information is kept confidential within a group, to prevent exploitation by others. While there exist theories that diversity of language can arise in order to distinguish ``us" versus ``them" \cite{haugen1972ecologyOfLanguage, trudgill1986dialectsInContact, mufwene2001ecologyOfLanguageEvolution, baker2003polari}, as well as several models \cite{abramsStrogatz2003modellingLanguageDeath, castello2006ordering, castello2009consensus, gong2008exploring, takahashi2007advancing, castello2013agent, dybiecsneppenmitarai2012culturalcenter},  the communicative aspect has typically been emphasized. However, considering the existence of diverse languages in the world and the frequent emergence of dialects by region, it is essential to also investigate the latter aspect.

This letter introduces a simple model that examines an antagonistic nature in agent interactions. Here we use the term ``language", even though the model abstracts all details in the language by focusing only on cooperative interaction by sharing codes and inhibitory one against commonality with others. With this abstraction, the model may also capture the dynamics of diversified communities with different culture or institutions, as seen in the  dynamics of nations \cite{turchin2018historical}, although we retain the term ``language" throughout. After introducing an evolutionary model with agents, we will show how communities diversify, where clusters with different fitness can coexist, with successive changes caused by the emergence of new communities  from the boundaries.


We define a total population size of $L\times L$ agents, placed on a two-dimensional grid with periodic boundary conditions \footnote{We also studied the behaviour of a mean-field model: See Supplement Section 1, Fig.1-3}. Each agent speaks a language $\vec{c}$, defined by a bitstring of length $B$ \cite{stauffer2007bitString}, for example
\begin{equation}
    \vec{c} = \underbrace{\left[0,0,1,0,1,1,0,0,1\right]}_B
\end{equation}
Every bit in the bitstring can be thought of as a ``feature" of a language, for example, the presence of tense, gender for inanimate objects, and so on. As our model is abstract, one could also consider a bit to represent a `word', a minimal unit that allows for communication. With the above definition, we could consider the mutual `understandability' of two languages as the number of times both the bitstrings have a $1$ in a particular position: if agent $A$ has a word for a concept while $B$ lacks it, communication is hindered. As such, it can be represented as a simple vector dot product of the two languages.

Each lattice site contains one agent, with their own language. Every time step, each agent $x$ has two types of interactions:
\begin{itemize} \itemsep0em
    \item[-] Local interactions: for the four nearest neighbors (up, down, left and right), fitness is gained from understanding each other:
    \begin{equation} \label{eq:alpha-fitness}
    {F_{x, y}}^\textrm{local} = \frac\alpha4 \, \cdot \mathcal{U}(\vec{c}_x,\vec{c}_y)
    \end{equation}
    \item[-] Global interactions: for every agent in the lattice, fitness is gained by having a different language from one another:
    \begin{equation}\label{eq:gamma-fitness}
     {F_{x,y}}^\textrm{global} = \frac{\gamma}{L^2} \, \cdot d_\mathcal{H}\left(\vec{c}_x, \vec{c}_y\right) 
    \end{equation}
\end{itemize}
where $\mathcal{U}(\vec{c}_x, \vec{c}_y) = \vec{c}_x \cdot \vec{c}_y = \sum\mathrm{AND}(\vec{c}_x, \vec{c}_y)$ represents the `understandability' and $d_\mathcal{H}\left(\vec{c}_x, \vec{c}_y\right) = \sum\mathrm{XOR}(\vec{c}_x, \vec{c}_y)$ represents the Hamming distance between the two languages.

\begin{figure*}
    \centering
    \includegraphics[width=\linewidth]{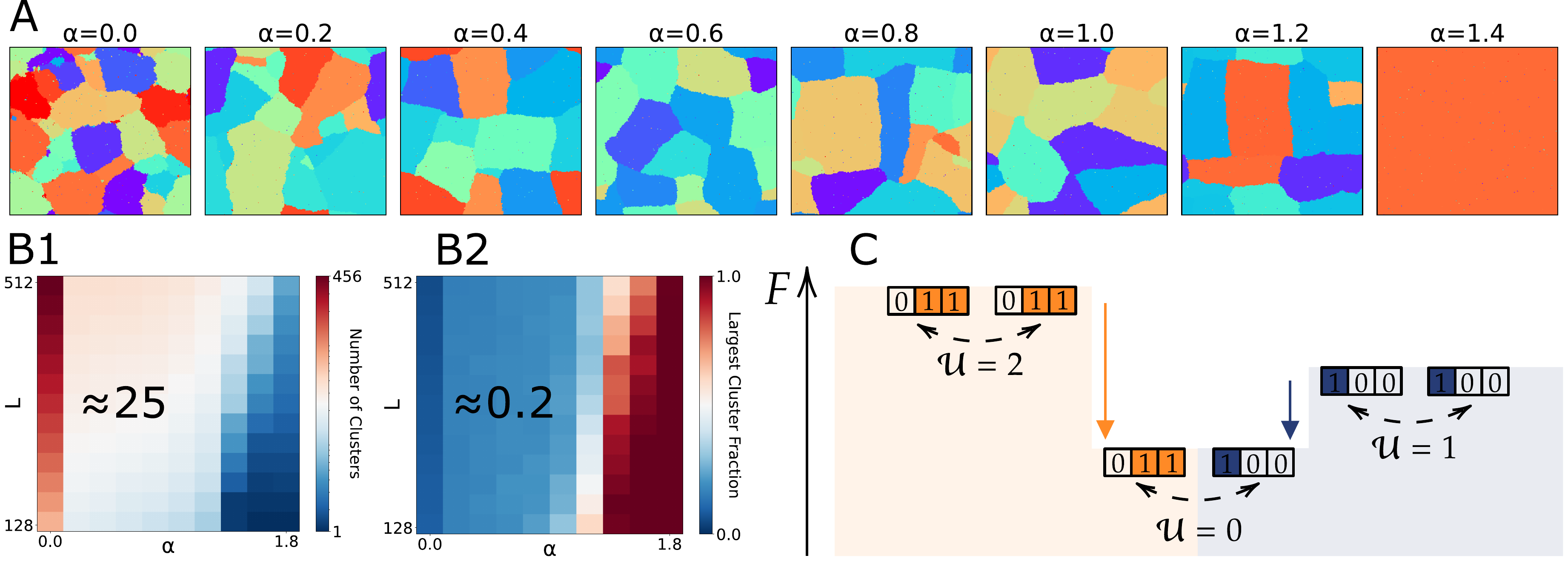}
    \caption{(A): Snapshots showing the state of the 2D lattice for varying alignment strength $\alpha$, for $\gamma$=1 $L$=256, $B$=16, $\mu$=0.001. Agents are colored randomly by their language. (B): Phase diagrams for the mean number of clusters \cite{Note3} (B1) and the largest cluster size divided by $L^2$ (B2) over $\alpha$ and $L$. We observe $\approx 20$-$30$ clusters, with the largest around $25\%$ of the system size robust to changing parameters, as long as $\alpha>0$ and is before the transition. (C): A schematic diagram of the boundary between two languages. The small rows represent the languages (011 and 100) and their heights represent their fitnesses. The $\alpha$ term of the fitness can be seen to drop, from 4 to 2 for the left language, and from 2 to 1 for the right language. This asymmetric dropoff allows languages with different bulk fitnesses to coexist, as long as their boundary fitnesses are equalized.}
    \label{fig:2D-lattice-raster}
\end{figure*}

In other words, the total fitness gained by an agent can be written as

\begin{equation} \label{eq:Fitness}
    F_x = \underbrace{\frac\alpha4 \,\sum_{y\in\mathrm{nbrs}}\mathcal{U}(\vec{c}_x,\vec{c}_y)}_\textrm{understandability} + \underbrace{\frac\gamma{L^2} \, \sum_{y=0}^{L^2}d_\mathcal{H}(\vec{c}_x,\vec{c}_y)}_\textrm{discommunication}
\end{equation}

According to the fitness, agents are replicated: We randomly select a site on the lattice, and check if its fitness is greater than any one of its four neighbors. If it is, it `invades' the weakest neighbor, replacing it with a clone of the `winner'. The newly created clone is immune to any successive invasion attempts this phase, and cannot be selected as a proposed replicator. We then continue this process until we have made $L^2/2$ trials for replication. Finally, after the reproduction is complete, we mutate each agent by modifying their bitstring, with each bit being flipped with an independent probability $\mu$. This makes up the next generation, and the evolutionary algorithm continues. As the selection depends only on relative fitnesses, we ensure that only the ratio of $\alpha/\gamma$ is a parameter, and as such $\gamma$ is set to unity. The system can be thought of as one with local activation (speaking easily understood languages) and global inhibition (speaking different language) \cite{reichardt2004detecting, reichardt2006statistical}, similar to a several pattern forming models \cite{turing1990chemical, bornholdt2001economicisingmodellocalglobalcoupling}.

We start all agents speaking the same ``unevolved" language consisting of a bitstring of $0$s, and perform the evolutionary simulation. Looking at the evolved state, we notice that agents form communities which speak the same language. We observe a transition into diversity of communities. While $\alpha/\gamma$ is large, speaking the same language is beneficial as all agents wish to speak a language which is easily understood. As such, we observe coarsening, and finally the entire system of agents speak one common language (a bit-string of only $1$s), with fluctuations corresponding to the mutation rate $\mu$. However, as $\alpha/\gamma$ is decreased, agents prefer to communicate in different languages, and the number of distinct languages increases.

Snapshots of the lattice for increasing $\alpha$ can be seen in \autoref{fig:2D-lattice-raster}A. We obtain diverse languages if $\alpha/\gamma \lessapprox 1.4$ \footnote{to be precise, the diversity transition is hard to quantify exactly. The position of the transition depends on $\mu$ as well as on $L$. The system also shows hysteresis (Supplement Fig.6) reminiscent of classical nucleation theory \cite{kalikmanov2012classical, xu2014nucleation}, which implies that the transition also depends on initial conditions. However, the transition is observed to be shallower (easier to homogenize) in the mean-field model and steeper (easier to diversify) in 1D. (Supplement Fig. 2)}. The clusters commonly have vertical or horizontal boundaries as a consequence of using a square lattice: straight lines are more robust to invasions than corners or diagonal lines. Clusters are also formed for $\alpha=0$, due to the local reproduction in the system. Interestingly, while different values of $\alpha/\gamma$ do not appear to significantly change the cluster size distribution \footnote{Due to the presence of mutations causing single random agents to have a different language, we ignore clusters of size 1.}, $\alpha=0$ is a special case, and appears to have a larger number of clusters, along with the presence of smaller clusters, than cases where $\alpha>0$ (\autoref{fig:2D-lattice-raster}B). The lattice size $L$ only weakly affects the number of clusters ($L$=128 leads to 20 clusters, while $L$=512 leads to 30: see supplement Fig. 5 for $L$, $B$ scaling) without changing the largest cluster size, due to the global inhibition ($\gamma$) term: `Winning' languages are penalized not when they reach a certain absolute size, but rather, when they reach a certain fraction of the system size.

One might expect that the coexisting clusters have the same fitness, but that is not the case. A peculiarity of the model is that the fitness on the boundaries are (nearly) all equally balanced. Fitness in this case can be considered analogous to pressure in bubble dynamics: if the fitness on a boundary is not equal, the stronger language can invade the weaker one. This, in turn, affects the $\gamma$ term of the fitness equation, causing the invader's fitness to drop, and the invaded's fitness to rise. This happens incrementally until all boundaries simultaneously equalize their fitnesses. It is important to note, however, that the \textit{bulk fitnesses} of languages are different: we can observe coexistence of strong species with weak ones, due to asymmetric fitness drop-offs at the boundaries (\autoref{fig:2D-lattice-raster}C).

\begin{figure}
    \centering
    \includegraphics[width=\linewidth]{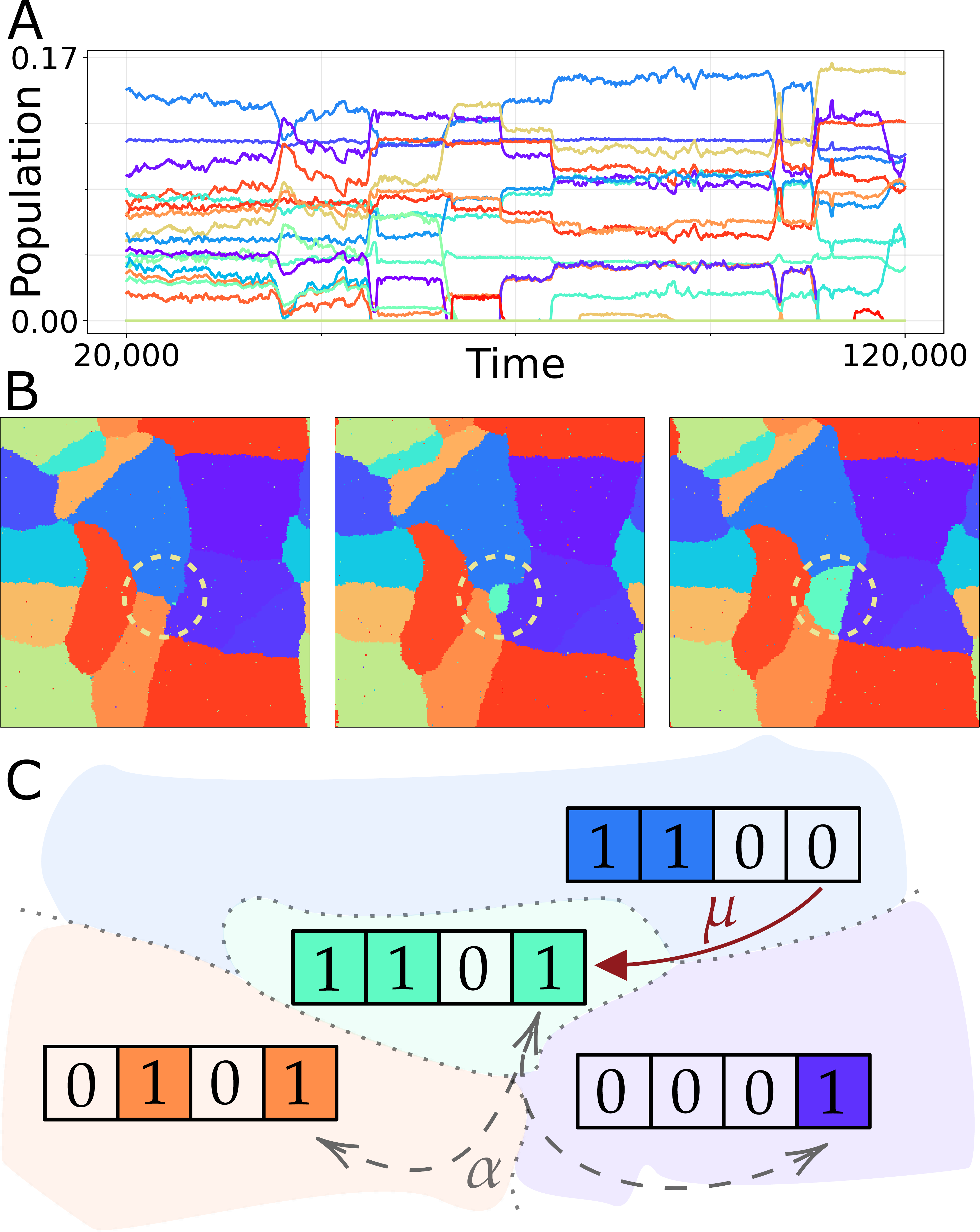}
    \caption{(A): The number of agents speaking a given language in the 2D lattice model, over time. For low $\mu$ ($\mu$=$10^-5$ in the figure), the system appears to stay in mostly `metastable' states, which can be destabilized in a quick transition. This transition moves the system to a new steady state, where populations remain roughly constant (reminiscent of punctuated equilibria \cite{bak1993punctuated, gould1993punctuated}). (B): Snapshots of the time evolution, showing the emergence of a new cluster. A mutation (cyan) forms on the boundary of three clusters, which proceeds to expand and `invade' its neighbors. Mutations in the bulk of a cluster are often unfit due to non-alignment with the neighbors, but cluster boundaries have lower fitness from alignment and thus are more vulnerable. (C): A schematic diagram for new community creation. Although the cyan language (1101) is created as a mutation from the blue (1100) in the top right, it contains `borrowed' features from the other languages, as that would increase the mutants fitness over one which lacks them (eg: if 1110 was mutated instead, it would have a lower fitness and promptly be eliminated).}
    \label{fig:2D-new-cluster-formation}
\end{figure}

Under the presence of mutations, the system never reaches a constant steady state, but appears to keep switching between `metastable' states, even at long time (\autoref{fig:2D-new-cluster-formation}A). This is due to the global inhibition introducing a context-dependent winner: While a language might do well in a certain environment, as it gets larger, it affects the global dynamics and promotes other languages based on the $\gamma$ term \footnote{Such transitionary dynamics over quasi-stationary states has been investigated by punctuated equilibria \cite{bak1993punctuated, gould1993punctuated}, ecosystem dynamics \cite{mallmin2024chaoticturnoverspecies} and high dimensional dynamical systems \cite{kaneko2003chaoticitinerancy}, though the mechanisms here are distinct from these.}. As such, there is no clear ``fittest" language in the system, and we cannot assign a potential to discuss a fitness landscape. At high mutation rate $\mu$, the frequency of switching increases to the point that we lose the metastable-switching behavior, and the dynamics resemble oscillatory fluctuations. Several communities appear to converge to similar sizes, a phenomenon we investigate in the 1D case.

Agents in the center of a cluster have a higher $\alpha$ fitness term than agents on the boundary, which disalign with some neighbors. This leads to the boundaries being less fit than the bulk. Snapshots of the lattice during one such `switching' event can be seen in \autoref{fig:2D-new-cluster-formation}B. While mutations happen everywhere in the system, mutants in the center of a cluster are likely to be removed due to the presence of several highly-fit neighbors. However, mutations on the boundary have a chance of having higher fitness than their neighbors and have a chance to grow, destabilizing the `steady state' of the system and pushing it to a new one.

As new languages appear through mutations, a phylogenetic tree of the evolutionary sequence can be depicted (Supplement Fig. 7). In our model, however, a mutation of language $c_x$ on the border with $c_y$ is unlikely to grow unless it aligns with $c_y$ (due to \autoref{eq:alpha-fitness}) and thus mutants that `borrow' from their neighbors are often selected (\autoref{fig:2D-new-cluster-formation}C). This is not represented in a phylogenetic tree, and is an aspect shared with real languages, which are reported to not solely be derived from a parent language, but also contain several `borrowed' features and words \cite{nelson2011networksborrowing, list2014networkslanguages}.

\begin{figure}
    \centering
    \includegraphics[width=\linewidth]{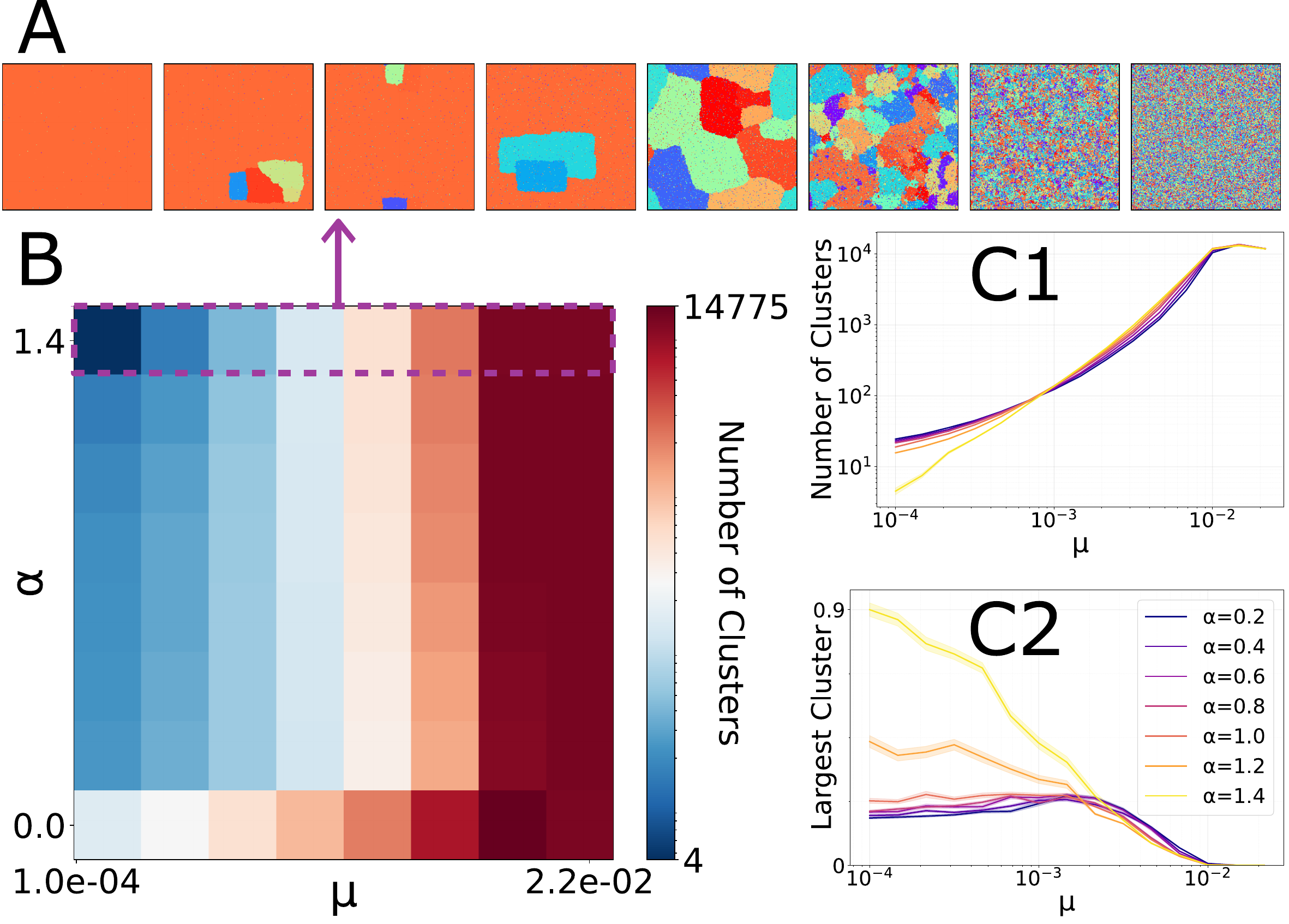}
    \caption{(A): Snapshots of a $L$=256 lattice with $B$=16 at $\alpha$=1.4 ($\gamma$=1), just above the transition point, showing the behavior for increasing mutation rate $\mu$. The homogeneous cluster can be destabilized by increasing $\mu$ (on a log scale for $\mu$), forming clusters that collapse into uncorrelated languages on further increasing $\mu$. (B): A phase diagram for the same parameters over alignment term $\alpha$ and mutation rate $\mu$ showing the number of clusters \cite{Note3} observed. While $\alpha$ does not appear to significantly change the number of clusters (similar to \autoref{fig:2D-lattice-raster}) increasing $\mu$ appears to increase the cluster number. (C): Plots of the number of clusters \cite{Note3} (C1) and the size of the largest cluster (C2) as a function of $\mu$. The scaling seems similar, robust to changes in $\alpha/\gamma$, and only weakly dependent on $L$ and $B$ (for $B$, assuming $\mu$ is scaled inversely: See supplement Fig.8,9).}
    \label{fig:2D-lattice-raster-mu}
\end{figure}

Next, we investigate the effect that the mutation rate $\mu$ has on the system. We can construct a similar phase diagram to the one in \autoref{fig:2D-lattice-raster}, except over $\alpha$ and $\mu$, for a fixed value of $L$ (\autoref{fig:2D-lattice-raster-mu}A,B). We observe that large mutation rate shrinks the cluster sizes, and is also able to destabilize a fully aligned cluster for the same ratio of $\alpha/\gamma$. We did not observe a strong dependence on the bitstring length $B$, as long as the mutation rate per bit $\mu$ is scaled inversely to preserve the probability of a mutation across the entire bitstring.

Increasing the mutation rate $\mu$ leads to an inability to effectively transfer information to the next generation, similar to the error catastrophe first discussed by Eigen \cite{eigen1979abstracthypercycle, eigen1988errorcatastrophe}. Unlike Eigen's original example, we lack an ``optimal" (fittest) language, as the best language to speak depends on your environment. However, the collapse of the structured state into an unstructured one allows us to draw parallels to the original idea. In addition, \autoref{fig:2D-lattice-raster-mu}C shows how the number of clusters and the largest cluster size scale with $\mu$. The behaviors for different $\alpha/\gamma$ appear to collapse on common scaling functions.


\begin{figure}
    \centering
    \includegraphics[width=\linewidth]{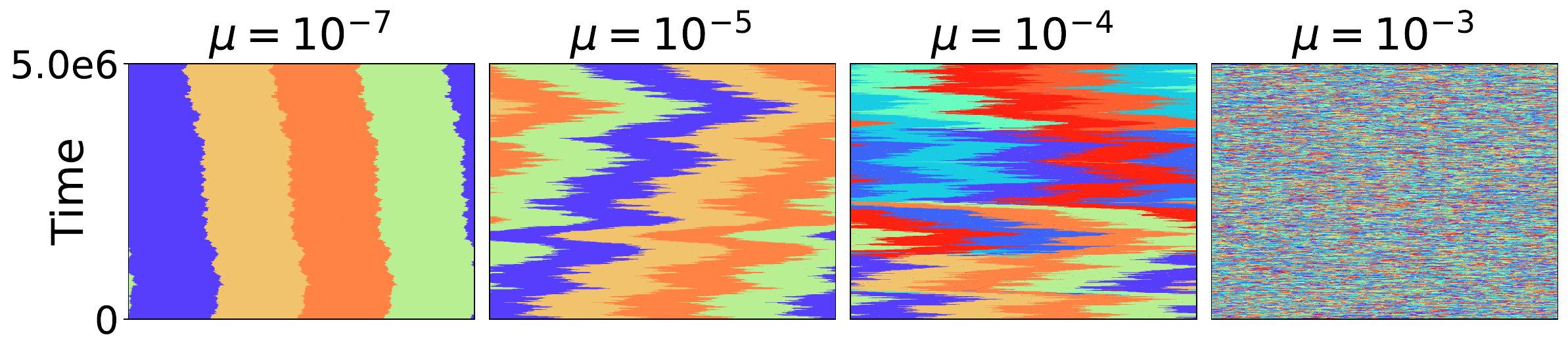}
    \caption{A plot of the 1D lattice over time, from the same initial condition, for different values of mutation rate $\mu$, for $L$=512, $B$=16. $\mu$=0 has stable boundaries which remain in place. Increasing $\mu$ leads to faster moving boundaries, the emergence of new langauges, and finally a collapse into an unstructured state.}
    \label{fig:1D-mutation-rate}
\end{figure}

Finally we investigate the behavior of our model on a one-dimensional (1D) lattice. Again, the clusters of different languages evolve, whose number increases with $\mu$ (\autoref{fig:1D-mutation-rate}). We observe that several clusters typically consist of the same number of agents. This can be explained in a simple case: consider that we have three different species, $c_0$, $c_1$, and $c_2$, with populations $N_0$, $N_1$ and $N_2$ respectively. At the boundary between $c_1$ and $c_2$, the total fitness must be equal, and thus we can derive (See supplement, Section II)
\begin{equation}
    \left[\frac\alpha\gamma - N_0\right](f_1 - f_2) = \left[d_\mathcal{H}(c_1, c_2)\right](N_1 - N_2)
\end{equation}
While there are several values of $f_1$, $f_2$, $N_1$ and $N_2$ that solve this equation, we typically observe that languages contain the same number of bits ($f_1 = f_2$), set by $\alpha/\gamma$. In that case, the only way to solve the equation is for $N_1$ to be equal to $N_2$, which can explain the alignment of species populations.

At low $\mu$, the switching behavior seen in \autoref{fig:2D-new-cluster-formation} is not observed. Instead, random mutations can cause shifts in the boundaries (Supplement Fig. 4). This can appear to behave as a traveling wave, however over very long time intervals it can be seen to be simply Brownian motion with inertia, where the source of the inertia comes from the global carrying capacity \footnote{However, in the specific case of a finite system with no parity between $\alpha$, $\gamma$, $L$ and $B$ (for instance,  $\alpha$=1, $\gamma$=1, $L$=31 and $B$=16), we can get into a situation where it is impossible for the fitness of all boundaries to be simultaneously equalized, which creates an unstable system. This unstable system can have traveling waves, which has stronger species invading weaker ones (as discussed in depth in ecosystem theory \cite{fisher1937wave}), but simultaneously allows a weaker species to consistently invade a stronger one due to the boundary imbalance and global carrying capacity, making the whole system analogous to a Penrose staircase.}.


Our model is abstract, as the complexities of language cannot be simply mapped to a 16-dimensional boolean vector. Furthermore, we assume that a single agent speaks only a single language, while bilingualism could be a powerful tool, allowing an agent to have a local `secret' language while still being able to communicate globally. Despite these issues, we feel that our toy model and its generality can explain diversity of language in a different way from traditional arguments.

In practice $\gamma$ and $\alpha$, which represent the `competitiveness' and `cooperativity' of an environment, are not fixed, but vary with time. They could also vary by sector: `languages' used in war are heavily encrypted and complex in order to prevent eavesdropping, while those used in trade are much more homogeneous.

If we consider an abstract definition of language, communication goes beyond just animals: microbes, for instance, communicate via exchange of chemicals in a phenomenon called quorum sensing \cite{waters2005quorum, whiteley2017progressquorum}. While there is plenty of evidence of an ever-evolving ``arms race" between microbes and phages (viruses that infect bacteria), we believe that there could be a ``cryptographical arms race" between microbial species, and perhaps even larger organisms like plants \cite{shrestha2020impact}.


The model shows switching between quasi-stable states, reminiscent of puctuated equilibria, which are discussed in language and cultural evolution \cite{kuteva1999languages, o2024punctuated}. In addition, the appearance of new languages on the boundaries is similar to existing theories for creole formation \cite{thomason2023languagecontactcreole, mufwene2001ecologyOfLanguageEvolution}, whereas the rise of new nations from the border is frequently observed in history \cite{muller2025shapingstatesintonations, turchin2018historical}.

In conclusion, much of the complex behavior in the model stems from two primary aspects: the presence of a high-dimensional state and the context-dependent fitness. Together, they allow for asymmetric fitness dropoffs at the boundaries and thus the coexistence of less-fit communities.


\textit{Acknowledgments:} We thank Kim Sneppen and Shunsuke Ichii for their helpful discussions. RFN and KK are supported by the Novo Nordisk Foundation Grant No. NNF21OC0065542. The data that support the findings of this article are openly available.

\nocite{letunic2024phylogenetictreeviz}

\bibliography{references}

\end{document}


\title{Supplementary information for \textit{A Self-Organized Tower of Babel: Diversification through Competition}}
\author{Riz Fernando Noronha \orcidlink{0009-0007-2923-3835}}
\author{Kunihiko Kaneko \orcidlink{0000-0001-6400-8587}}
\affiliation{Niels Bohr Institute, University of Copenhagen, Copenhagen 2200, Denmark}

\maketitle

\onecolumngrid

\section{Mean-Field evolutionary model}

Similar to the lattice model described in the main text, we create a mean-field model with a total population size of $N$ agents. Once again, each agent speaks a language $\vec{c}$, defined by a bitstring of length $B$.

In each generation, every agent plays a `match' with \textbf{every other} agent, and together they gain fitness. The fitness of an agent $x$ in a single time-step is:
\begin{equation} \label{eq:Fitness}
    \mathcal{F_x} = \sum_{y=0}^N \left( \underbrace{\gamma \,d_\mathcal{H}(\vec{c}_x,\vec{c}_y)}_\textrm{discommunication} + \underbrace{\alpha \,\mathcal{U}(\vec{c}_x,\vec{c}_y)}_\textrm{understandability} \right)
\end{equation}

At the end of a generation, the average fitness accumulated by an agent is calculated, and the top 50\% of agents are chosen to reproduce. Each agent creates two copies of itself, and every bit in the child's bitstring mutates (flips) with probability $\mu$. Then, the children make up the new generation, and the algorithm continues.

\begin{figure}
    \centering
    \includegraphics[width=0.49\linewidth]{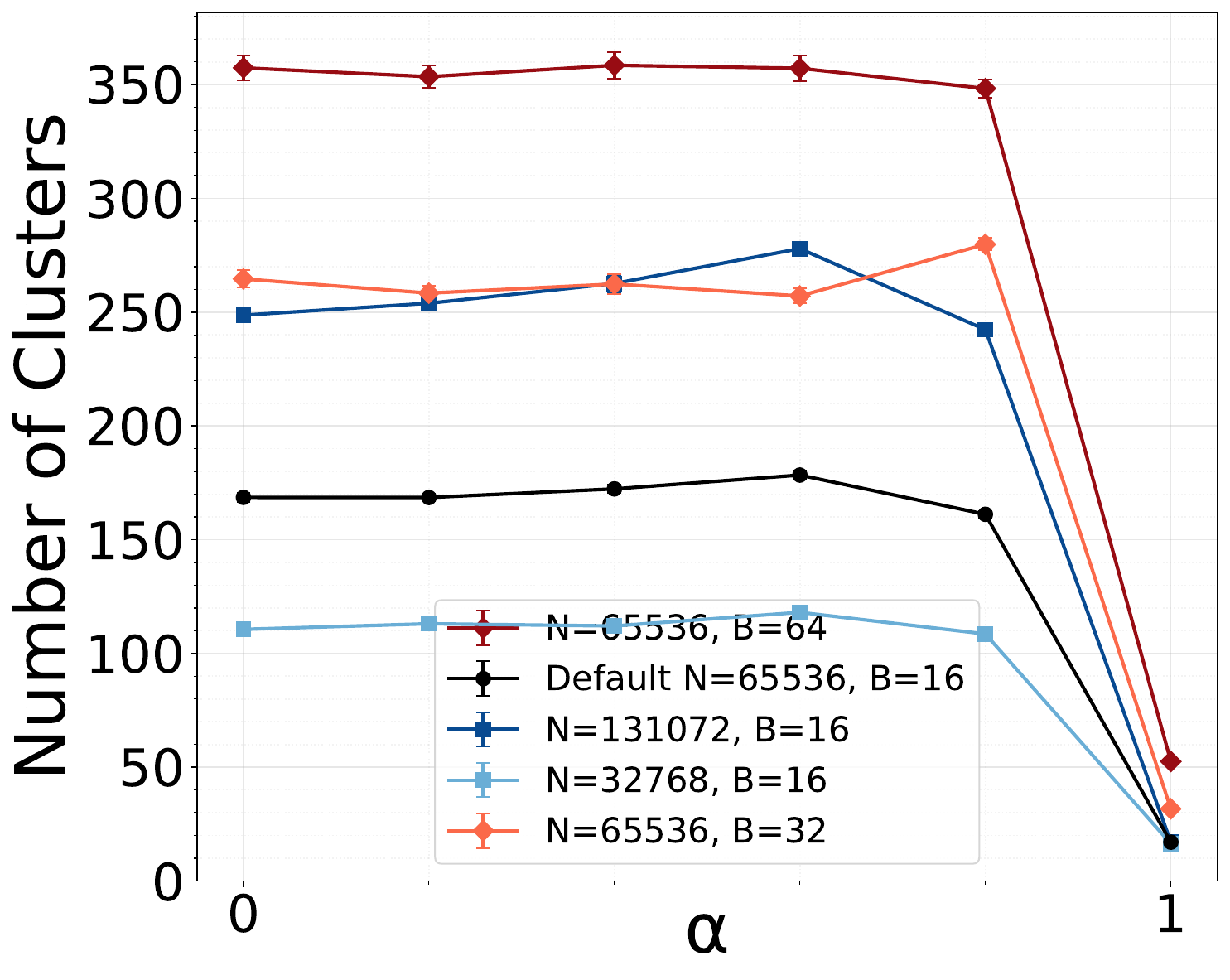}
    \includegraphics[width=0.49\linewidth]{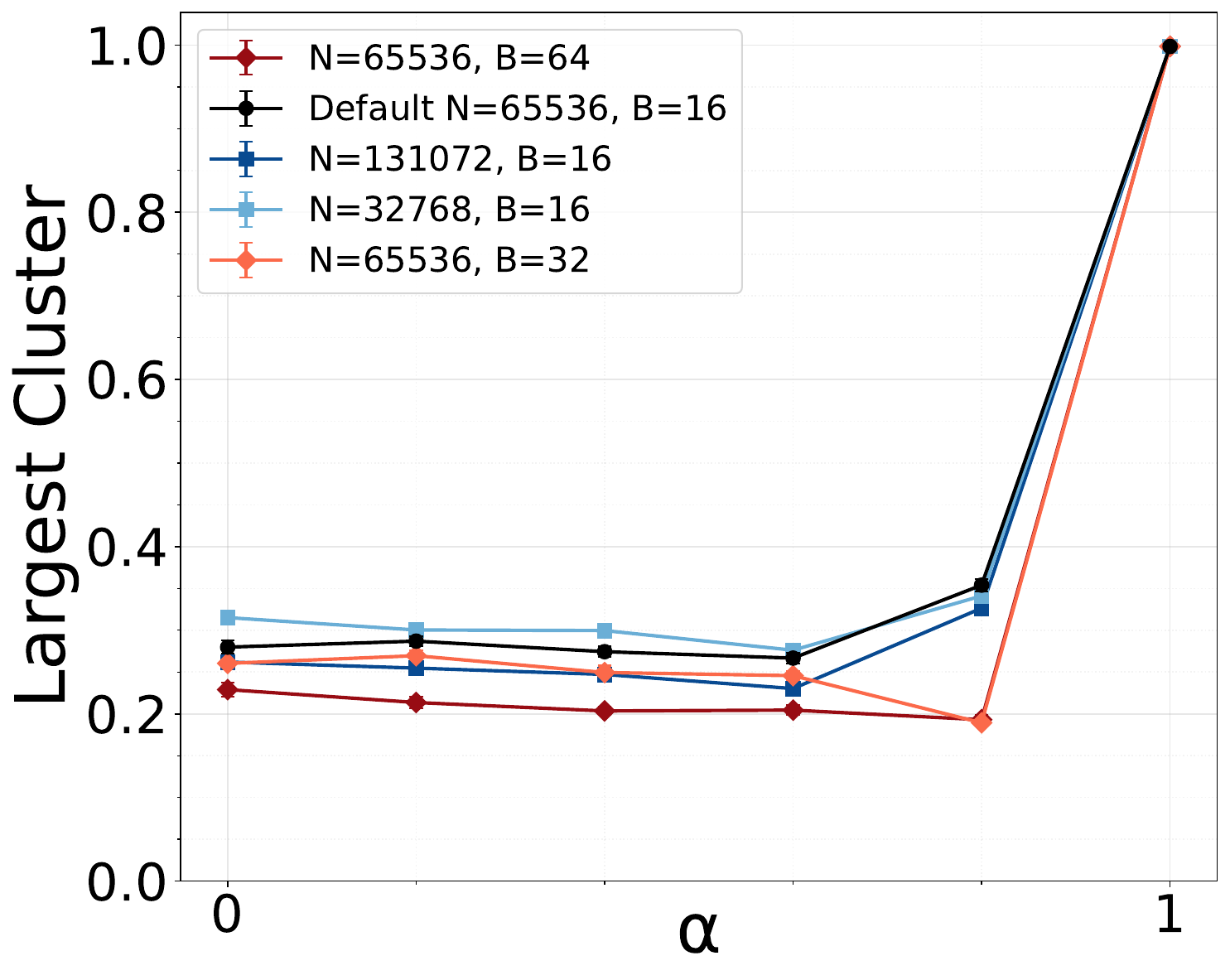}
    \caption{A plot of the number of languages excluding languages of size 1 (left) and the largest language's size (right) against alignment strength $\alpha$, for a system with $\gamma$=1, $\mu$=0.0001. Different lines indicate results upon changing $L$ (blue) or $B$ (red). The dominant language's size appears to be somewhat independent of $L$ and $B$, though larger systems or bitstrings and system sizes can allow for a larger diversity of languages. Similar to the 2D case, a transition can be observed at $\alpha\approx\gamma$, however, the transition occurs quicker, i.e, $\alpha<\gamma$, for instance. Note that the cluster number is significantly different with different system sizes, unlike the 2D case \autoref{fig:2D-lines-alpha-scaling}.}
    \label{fig:meanfield-lines-alpha-scaling}
\end{figure}

We start all agents speaking the same ``unevolved" language consisting of a bitstring of $0$s, and perform the evolutionary simulation, as in the 2D case. We once again observe a transition into the diversity of languages (\autoref{fig:meanfield-lines-alpha-scaling}). The transition is observed to happen earlier. This can be understood by considering the behavior of each of the two terms in the fitness equation:
\begin{enumerate}
    \item The $\alpha$ term (alignment) occurs only locally, and thus acts on the dimensionality of the space the simulation is embedded in.
    \item The $\gamma$ term (differentiation) occurs globally, regardless of the dimensionality of the space, and thus is similar to a mean-field (infinite dimensional) interaction.
\end{enumerate}

In the letter, we primarily talk about the 2D model, and a recurring theme of a tradeoff between $\alpha$ and $\gamma$ arises: namely, the need to align, and the need to differentiate. When we consider the scaling across dimensions, we once again come across the competition of the $\alpha$ and $\gamma$ terms but this time for a different reason. Now, one should view it as the tradeoff between a low-dimensional interaction (alignment) and a high dimensional densely connected one (differentiation). Due to this, we should expect the observed phenomena to get more and more extreme as we lower the dimension, as we increase the separation of `dimensional scales' further.

\begin{figure}
    \centering
    \includegraphics[width=0.49\linewidth]{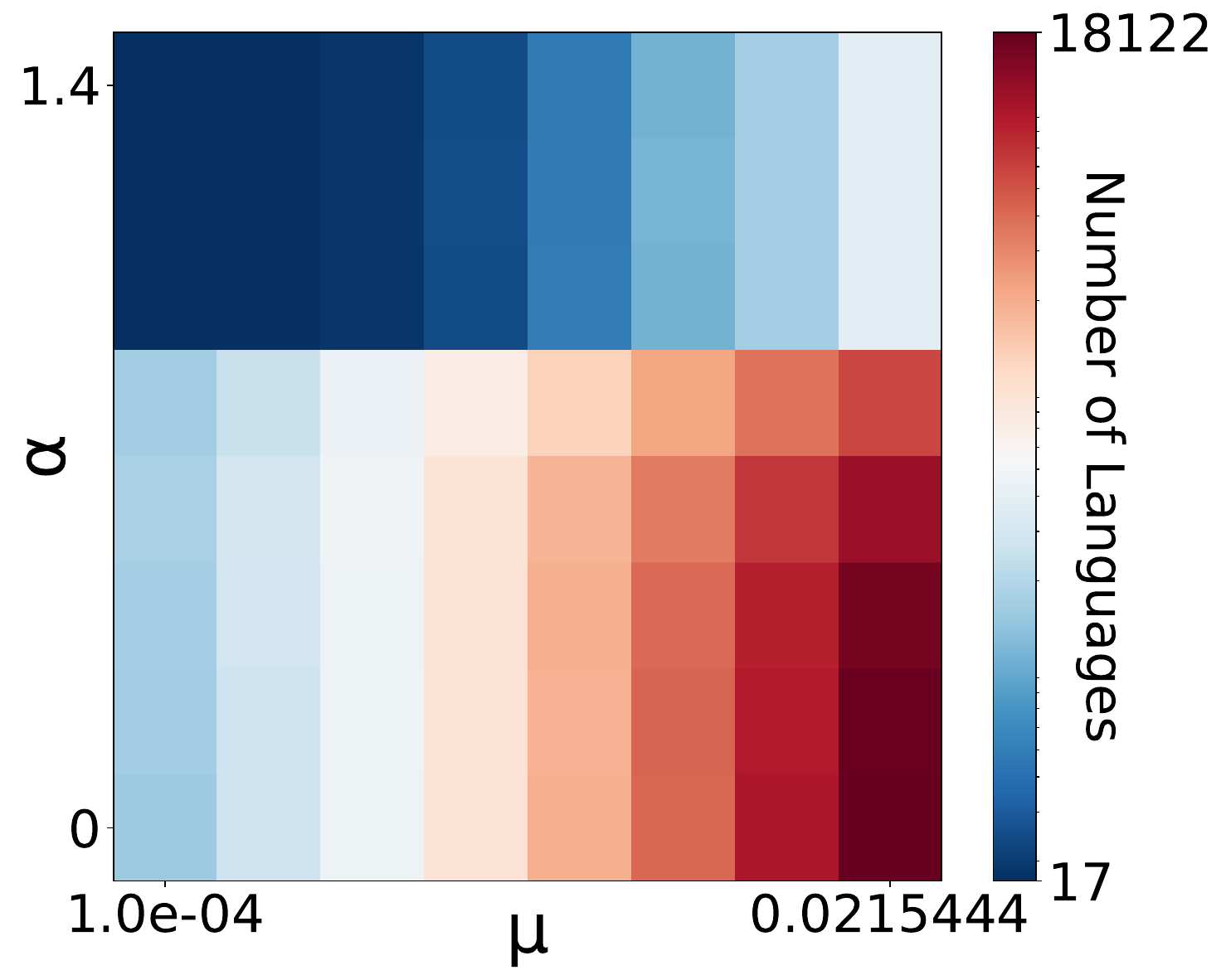}
    \includegraphics[width=0.49\linewidth]{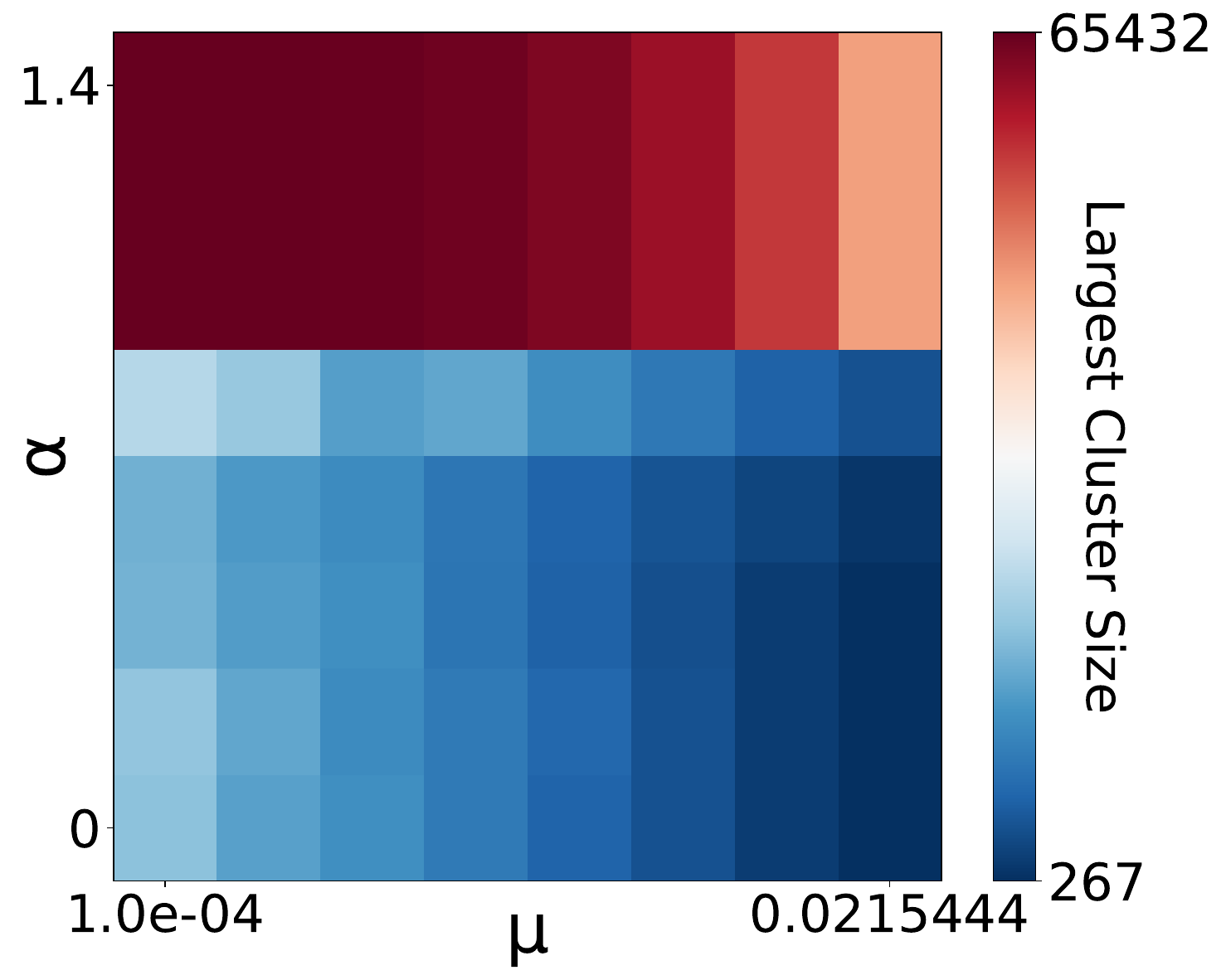}
    \caption{A heatmap of the number of languages with more than 1 speaker (left) and the largest cluster size (right) against alignment strength $\alpha$ and differentiation pressure $\gamma$, for a system with $N$=65536, $B$=16, $\gamma$=1. The axes are the same as in Fig. 3A, but the earlier transition can be seen easily. Once again, mutation rate causes an increase in the number of languages. However, it appears that the `breaking up of a majority cluster' just above the transition isn't seen in the mean-field case, as the parameters far away from the transition display similar behaviour.}
    \label{fig:meanfield-raster-alpha-mu}
\end{figure}

\begin{figure}
    \centering
    \includegraphics[width=0.5\linewidth]{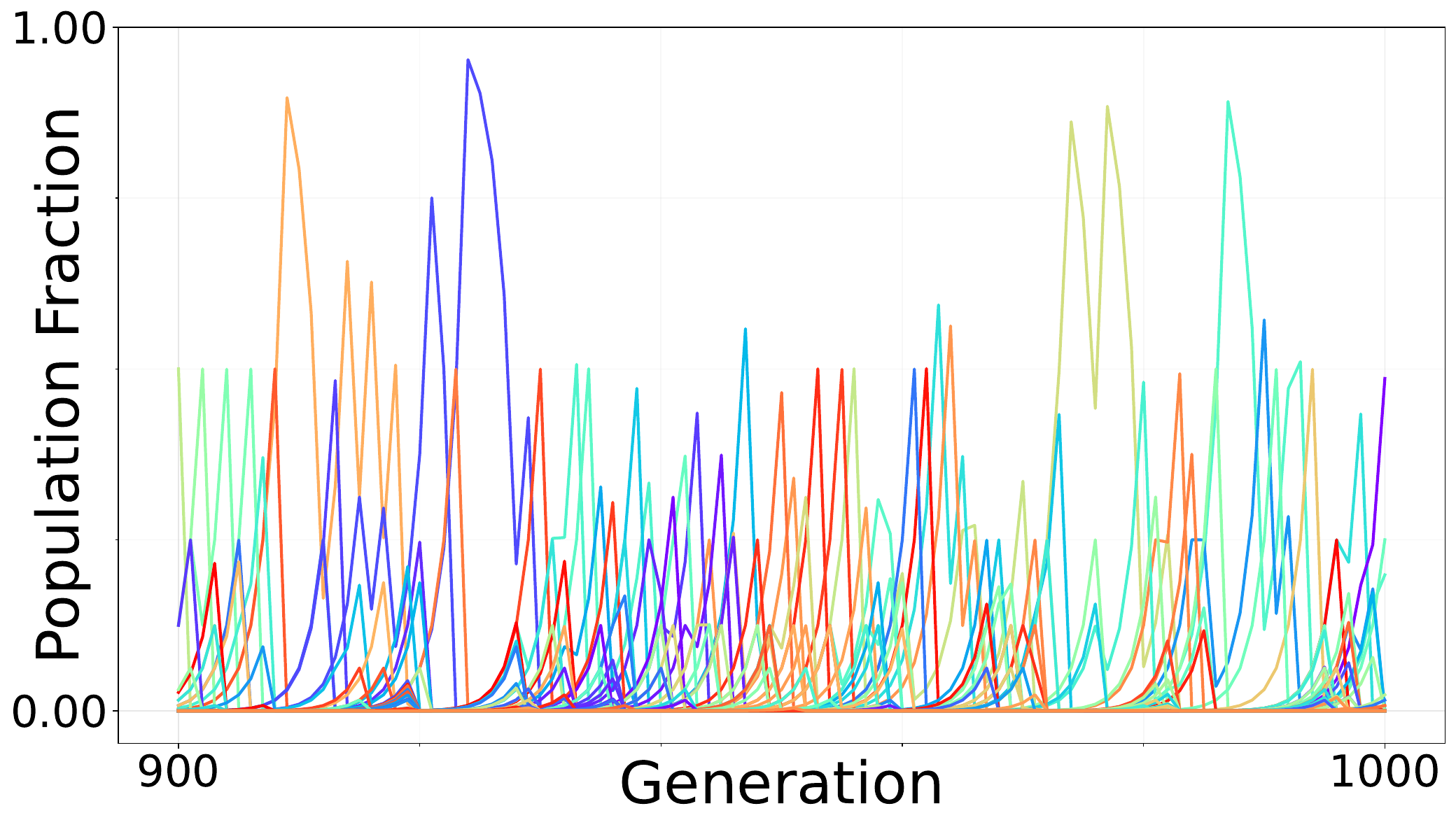}
    \caption{A timeseries of the populations of speakers of each language in the mean-field model, with $N$=65536, $B$=16, $\alpha$=0.4, $\gamma$=1, $\mu$=$10^{-5}$. Extremely fast switching is observed, where at a moment in time, only a single language dominates, but collapses with a few generations as a mutant takes over (on larger systems, due to the presence of more mutants, switching happens even faster, and a species).}
    \label{fig:super-fast-switching-mf}
\end{figure}

Thus, the mean-field model is, in many ways, a more extreme version of the two-dimensional model. For instance, in 1D, the evolved populations appeared relatively stable, while in 2D we had switching events when a mutant took over. \autoref{fig:super-fast-switching-mf} shows how the mean-field case has extremely fast switching, as a mutant has a large $\gamma$ term and can quickly dominate. In addition, the phase boundary for $\alpha/\gamma$ between alignment and diversity increases beyond unity as we change from the mean-field, to the 2-dimensional, and finally to the 1-dimensional model.

\clearpage

\section{1-Dimensional Evolutionary Model}

\subsection*{Deriving an equation based on boundary fitness being equalized}

We have $M$ languages: $c_1$, $c_2$, and $c_3 \dots c_{M}$. A language $c_i$ has a population (number of speakers) $N_i$, and the bitstring consists of $f_i$ ones. Without loss of generality, we consider the boundary between $c_1$ and $c_2$. If the fitness at the boundary is equal, then,

\begin{align*}
    \textrm{Global fitness of 1} + \textrm{Local fitness of 1} = \textrm{Global fitness of 2} + \textrm{Local fitness of 2} \\
    \gamma \left[ N_1 d_\mathcal{H}(c_1, c_1) + N_2 d_\mathcal{H}(c_1, c_2) + \sum_{i=3}^N N_i d_\mathcal{H}(c_1, c_i) \right] + \alpha\left[U(c_1,c_1) + U(c_1, c_2)\right] \\
    = \gamma \left[ N_1 d_\mathcal{H}(c_2, c_1) + N_2 d_\mathcal{H}(c_2, c_2) + \sum_{i=3}^N N_i d_\mathcal{H}(c_2, c_i) \right] + \alpha\left[U(c_2,c_1) + U(c_2, c_2)\right] \\
\end{align*}

Using the relations:
\begin{align*}
    U(c_i, c_i) = f_i \\
    d_\mathcal{H}(c_i, c_i) = 0
\end{align*}

\begin{align*}
    \gamma \left[ N_2 d_\mathcal{H}(c_1, c_2) + \sum_{i=3}^N N_i d_\mathcal{H}(c_1, c_i) \right] + \alpha\left[f_1 + U(c_1, c_2)\right] 
    = \gamma \left[ N_1 d_\mathcal{H}(c_2, c_1) + \sum_{i=3}^N N_i d_\mathcal{H}(c_2, c_i) \right] + \alpha\left[U(c_2,c_1) + f_2\right] \\
    \alpha\left[f_1 + U(c_1, c_2) - U(c_2,c_1) - f_2\right] 
    = \gamma \left[ N_1 d_\mathcal{H}(c_2, c_1)+ \sum_{i=3}^N N_i d_\mathcal{H}(c_2, c_i) - N_2 d_\mathcal{H}(c_1, c_2) - \sum_{i=3}^N N_i d_\mathcal{H}(c_1, c_i) \right]
\end{align*}

\begin{equation}
\frac\alpha\gamma\left[f_1 - f_2\right] = d_\mathcal{H}(c_2, c_1) \left[N_1 - N_2\right]+ \sum_{i=3}^N N_i \left[d_\mathcal{H}(c_2, c_i) - d_\mathcal{H}(c_1, c_i)\right] 
\end{equation}

Let us consider a simplified case where $M=3$, and the last language $c_3$ consists of only $1$s. Then, $d_\mathcal{H}(c_i, c_3) = B - f_i$

\begin{align*}
\frac\alpha\gamma\left[f_1 - f_2\right] = d_\mathcal{H}(c_2, c_1) \left[N_1 - N_2\right]+ N_3 \left[d_\mathcal{H}(c_2, c_3) - d_\mathcal{H}(c_1, c_3)\right]      \\
\frac\alpha\gamma\left[f_1 - f_2\right] = d_\mathcal{H}(c_2, c_1) \left[N_1 - N_2\right]+ N_3 \left[(B - f_2) - (B - f_1)\right]      \\
\end{align*}

\begin{equation}
\left(\frac\alpha\gamma - N_3 \right)\left[f_1 - f_2\right] = d_\mathcal{H}(c_2, c_1) \left[N_1 - N_2\right]
\end{equation}


\begin{figure}
    \centering
    \includegraphics[width=0.8\linewidth]{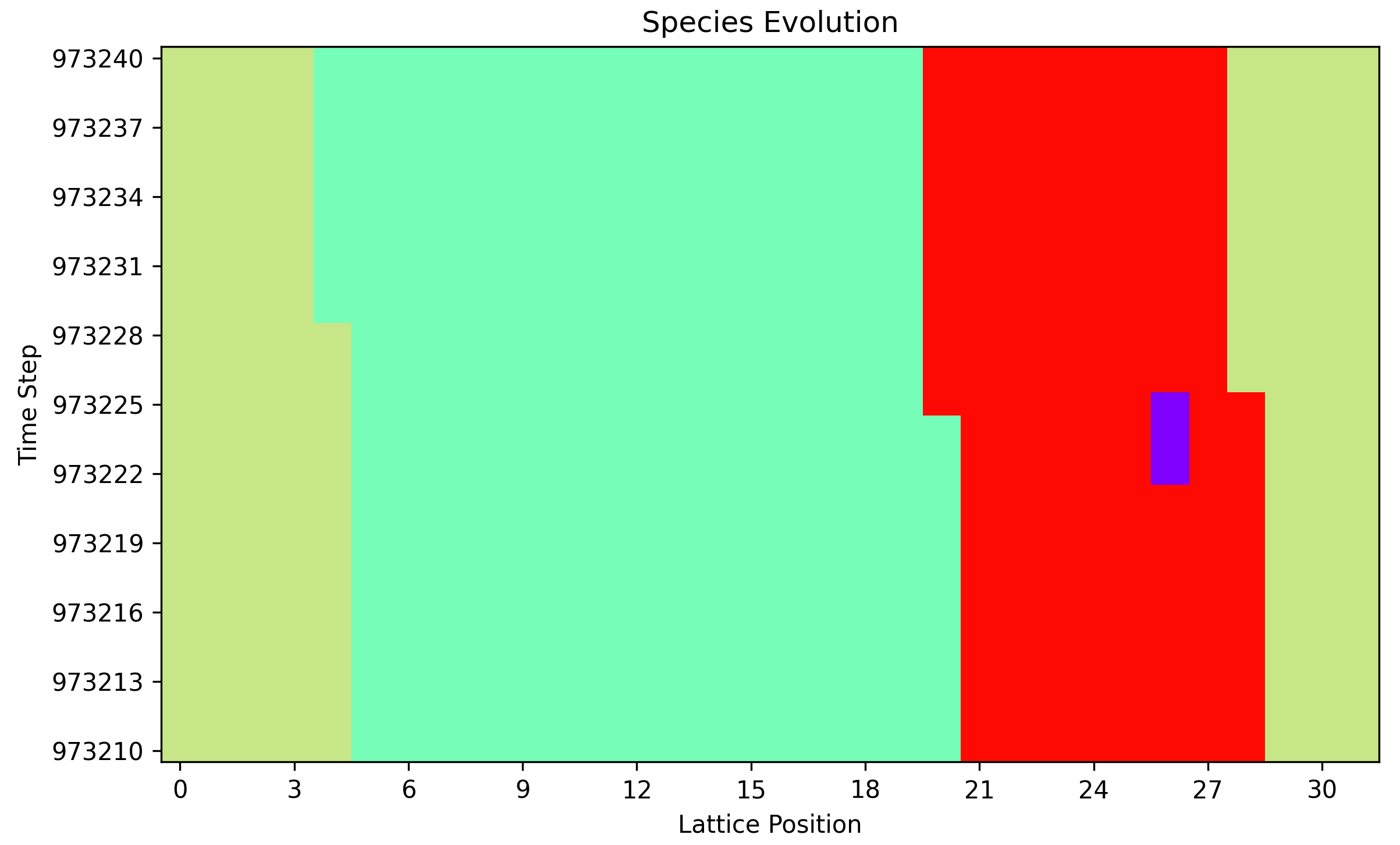}
    \caption{A diagram of the 1-dimensional lattice, showing how mutation, together with stochastic reproduction, can induce a shift in the boundaries. First, a purple mutant forms in the red region, which strengthens red, as it's $\gamma$ fitness increases. Red then proceeds (stochastically) to invade to the left before removing the mutant, which causes a chain reaction. When the mutant is finally removed, red is weaker (as it is larger due to the previous invasion), and can be invaded from the right, leading to the entire boundary being shifted.}
    \label{fig:1D-mutation-boundary-shift}
\end{figure}

\clearpage

\section{2-Dimensional Evolutionary Model}


\begin{figure}[h]
    \centering
    \includegraphics[width=0.49\linewidth]{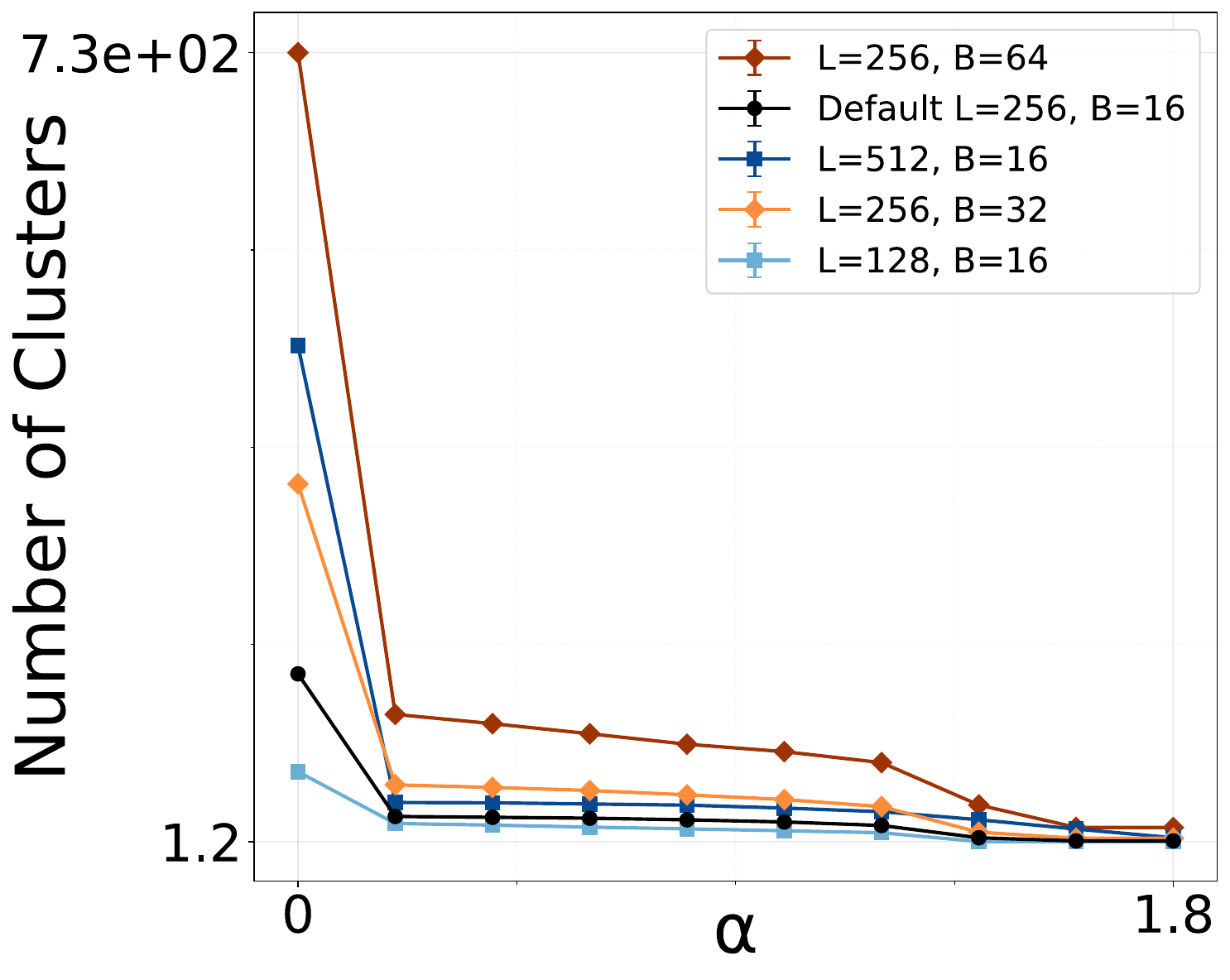}
    \includegraphics[width=0.49\linewidth]{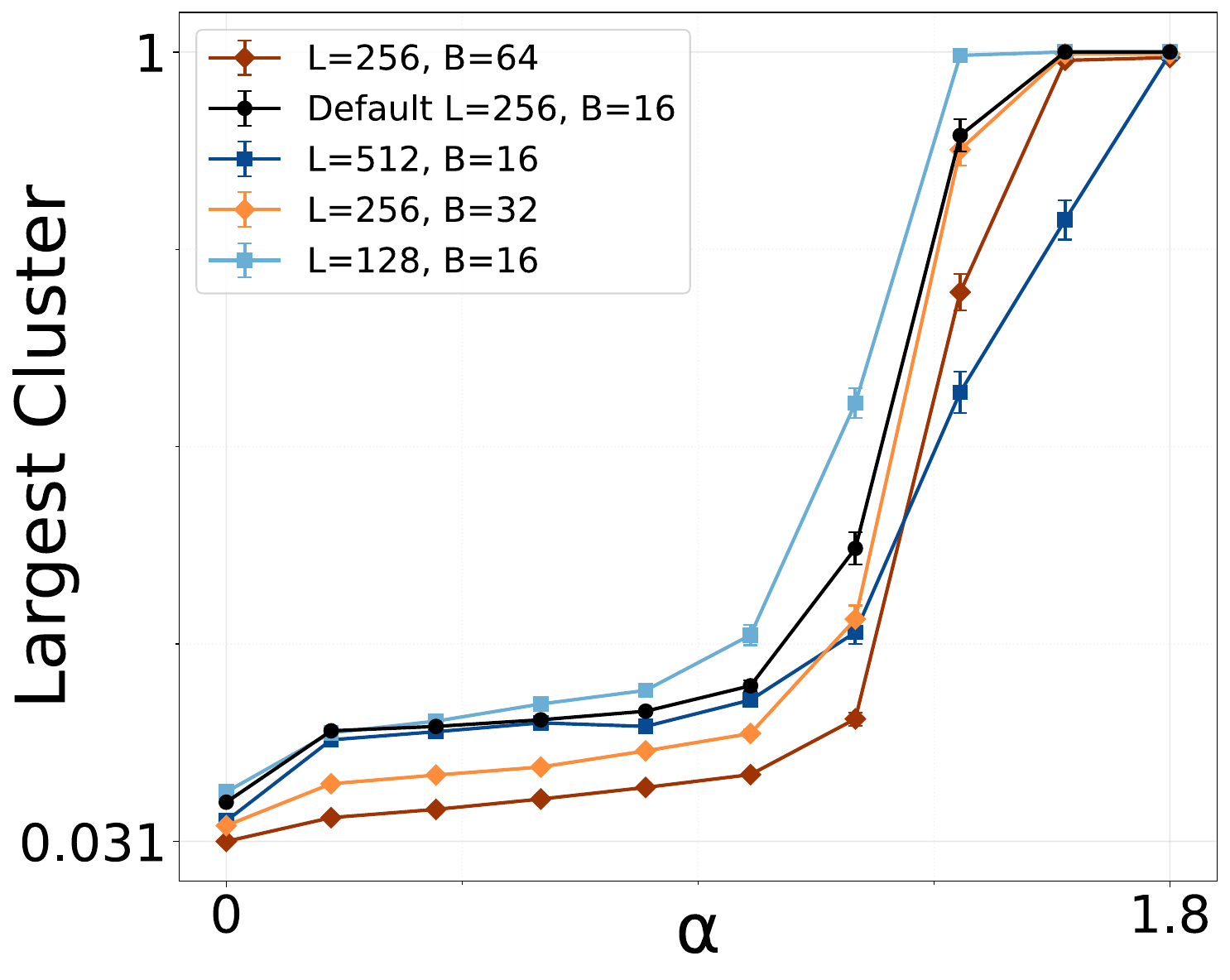}
    \caption{Similar to \autoref{fig:meanfield-lines-alpha-scaling}, a plot of the number of languages with more than 1 speaker (left) and the largest language's size (right) against alignment strength $\alpha$, for a system with $\gamma$=1, $\mu$=0.0001. Different lines indicate results upon changing $L$ (blue) or $B$ (red, though $\mu$ is scaled inversely). Both metrics indicate that the behavior of the system does not significantly change by modifying the system size, although longer bitstrings appear to decrease the size of the largest cluster and also increase the number of clusters. \autoref{fig:2D-lines-alpha-scaling}.}
    \label{fig:2D-lines-alpha-scaling}
\end{figure}


\begin{figure}[h]
    \centering
    \includegraphics[width=0.49\linewidth]{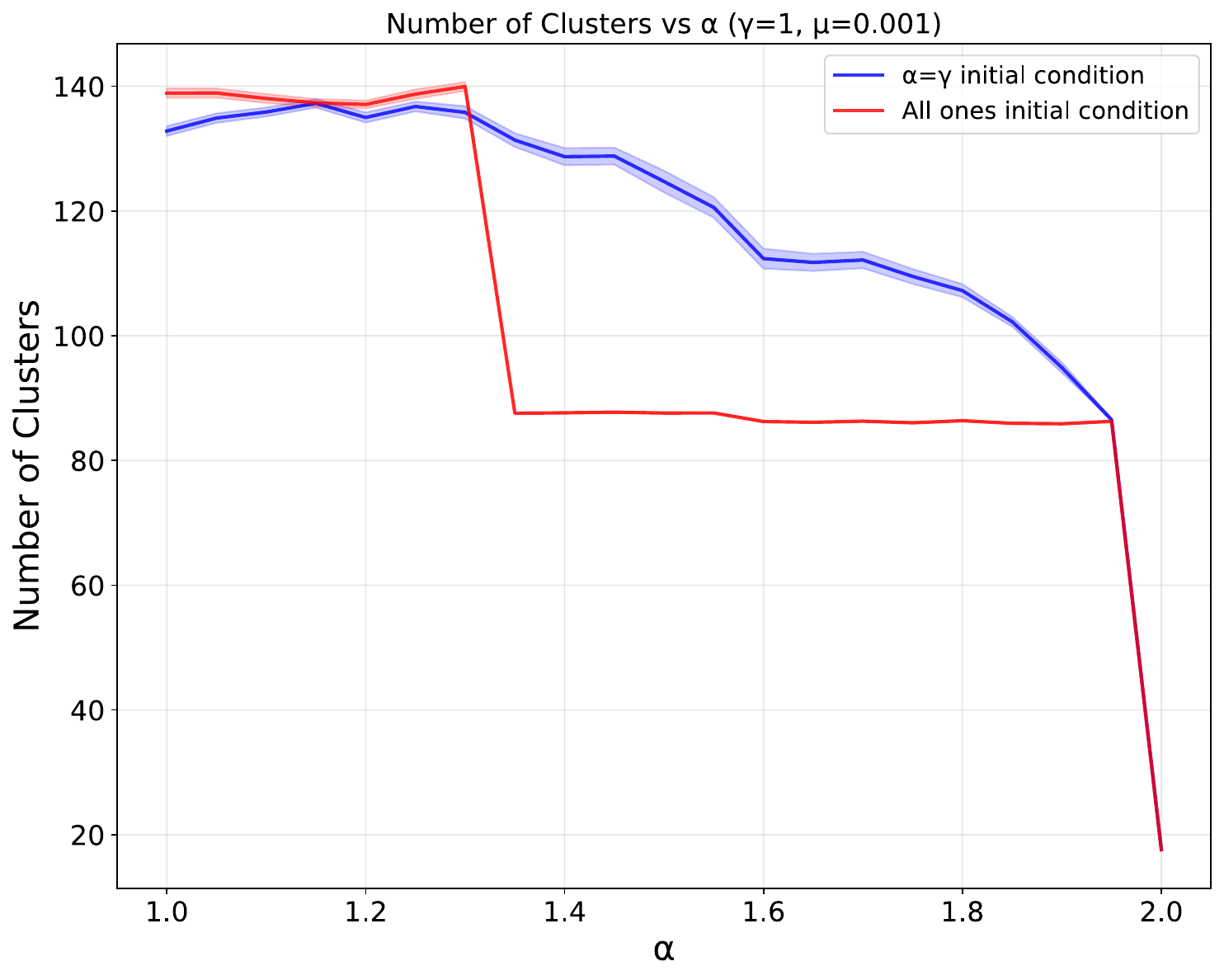}
    \includegraphics[width=0.49\linewidth]{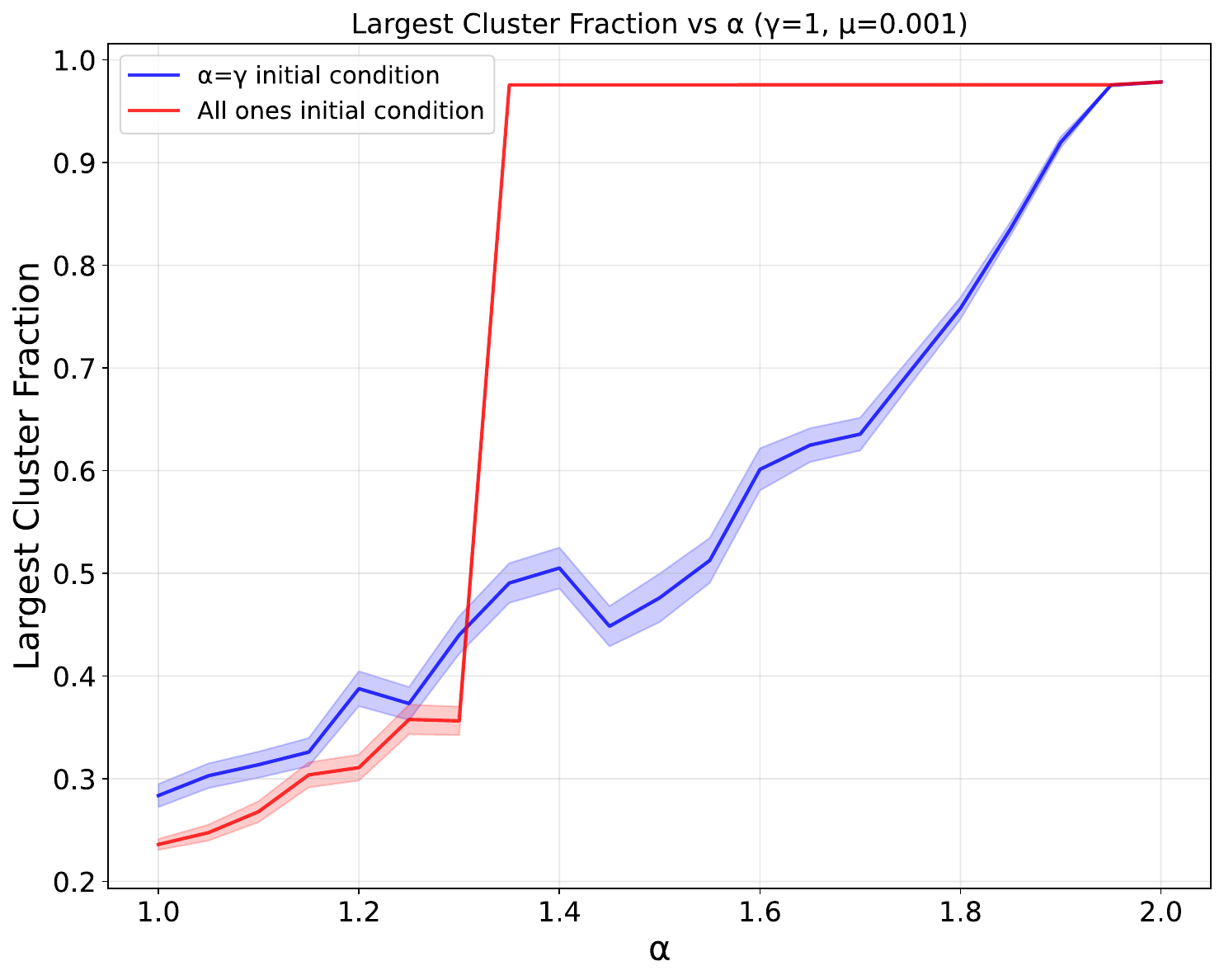}
    \caption{The system can exhibit hysteresis near the transition: a plot of the number of clusters (left) and largest cluster size (right) versus $\alpha$, for $L$=256, $B$=16, $\gamma$=1, $\mu$=0.001 for two different initial conditions. The red line represents an initial condition of all agents speaking the same language of a bitstring of 1s, while the blue line starts from an evolved state for $\alpha$=1, with diversity. If the initial conditions contain clusters, those clusters are robust and can survive, while if the initial condition contains one large cluster, diversity is rare as mutants beyond some critical size are needed to expand as in classical nucleation theory [39, 40].}
    \label{fig:largest-cluster-fraction}
\end{figure}


\begin{figure}
    \centering
    \includegraphics[width=\linewidth]{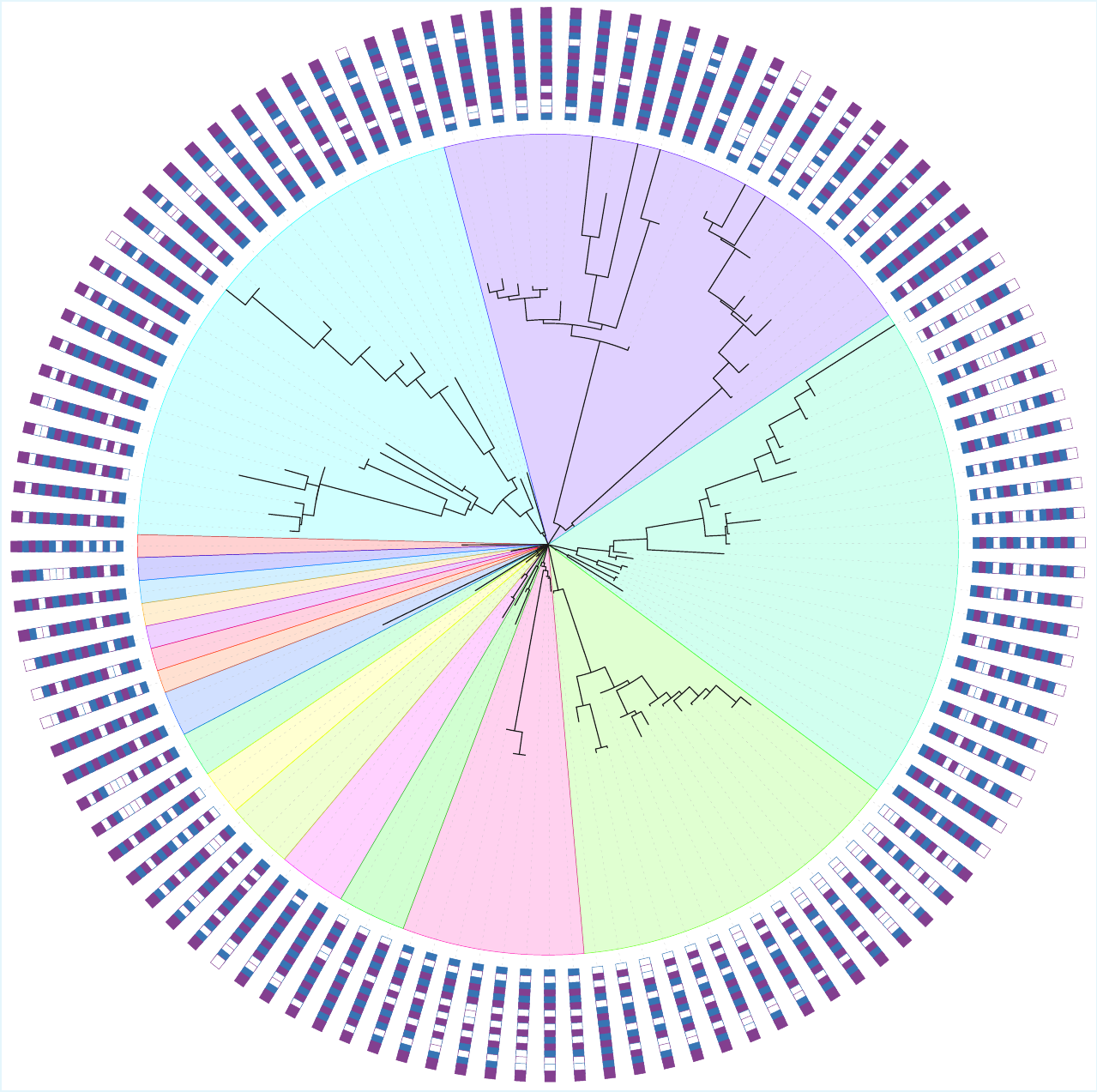}
    \caption{A phylogenetic tree showing the evolutionary path observed in the 2D model. The radial coordinate indicates time, and the branches show new languages created through mutations. Colors indicate a clade of languages originating from a common ancestor. The bitstring pattern is shown on the outside. Parameters are $L$=$256$, $\gamma$=1, $\alpha$=1, $B$=16, $\mu$=0.001. Created with iTOL [38]. Note, however, that mutants at the boundary can expand when they mimic the patterns of the neighboring domain, which can be regarded as an `emergent' horizontal gene transfer (Fig. 2C). Thus, it is not the most accurate way of depicting the evolutionary sequence.}
    \label{fig:phylogenetic-tree}
\end{figure}


\begin{figure}
    \centering
    \includegraphics[width=0.5\linewidth]{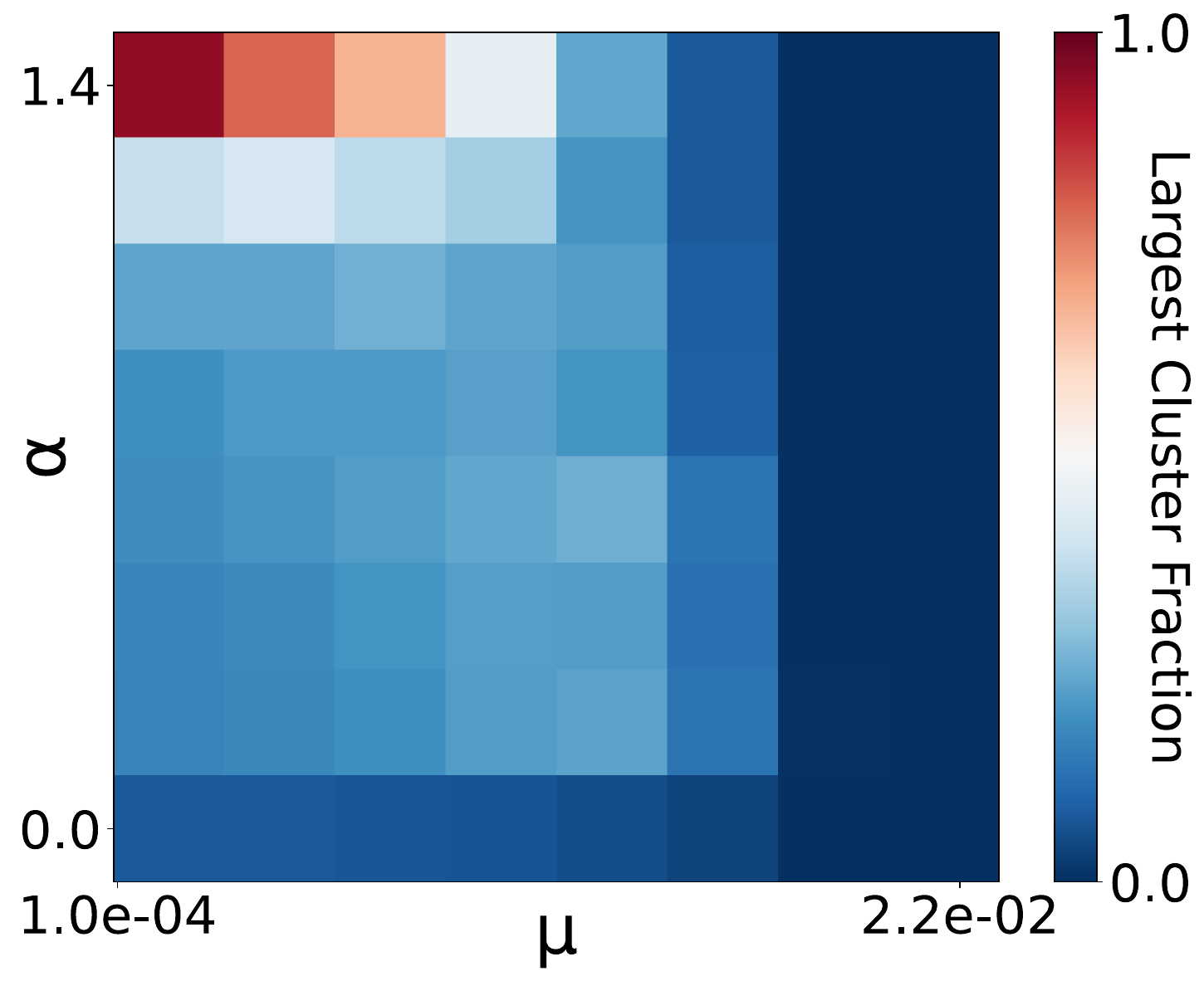}
    \caption{The size of the largest cluster as a function of the alignment strength $\alpha$ and the mutation rate $\mu$ (similar to Fig.3B from the letter), for a system with $L$=256, $B$=16, $\gamma$=1. Unlike the number of clusters, the largest cluster fraction appears to not significantly change until reaching near the transition (also see the right figure of \autoref{fig:mu-scaling-LB}). An exception can be seen in the top row, where low mutation rates mean the system is past the transition, and only a single cluster exists (which can be destabilized with reproductive noise).}
    \label{fig:alpha-mu-raster-largest-cluster}
\end{figure}

\begin{figure}
    \centering
    \includegraphics[width=0.32\linewidth]{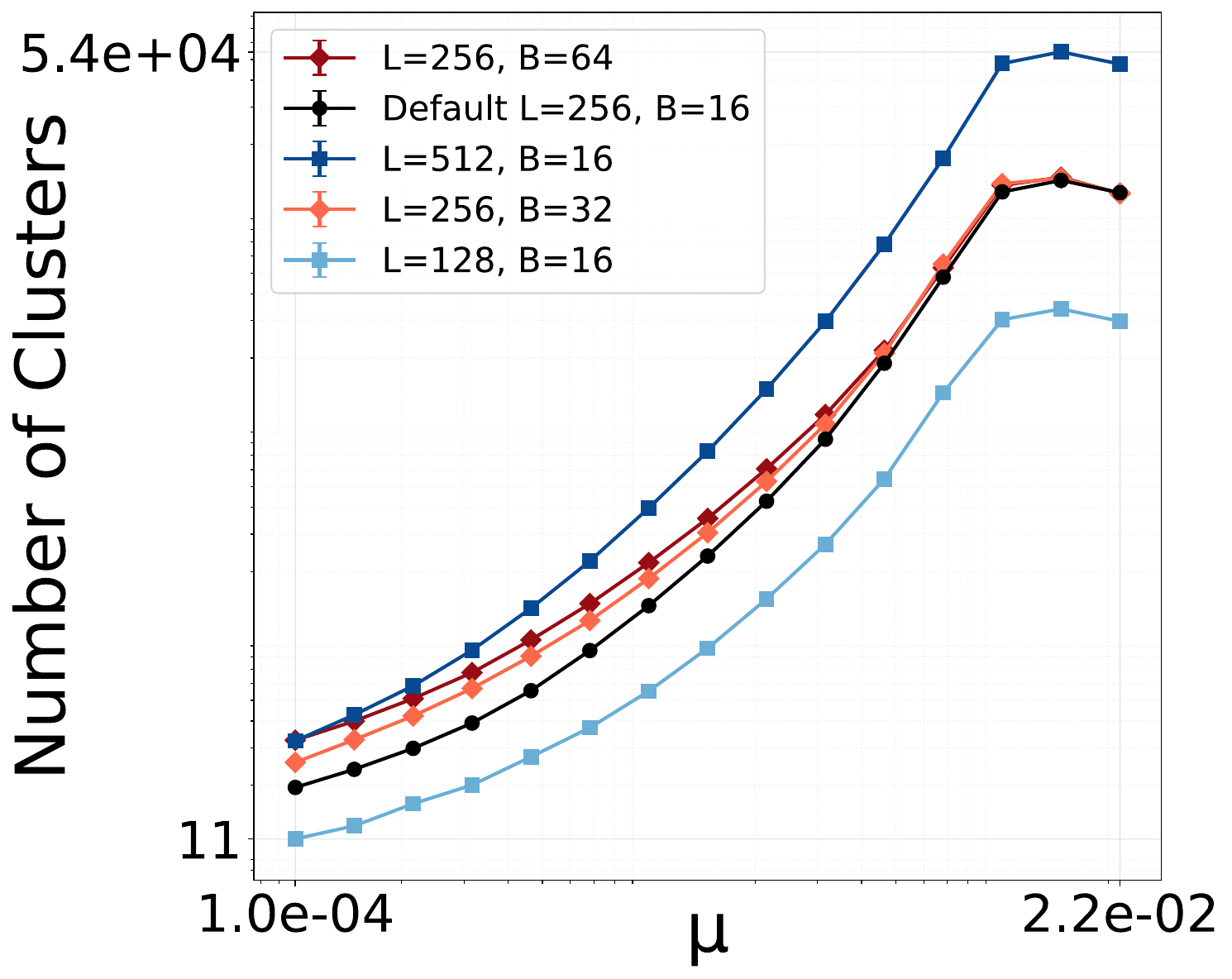}
    \includegraphics[width=0.32\linewidth]{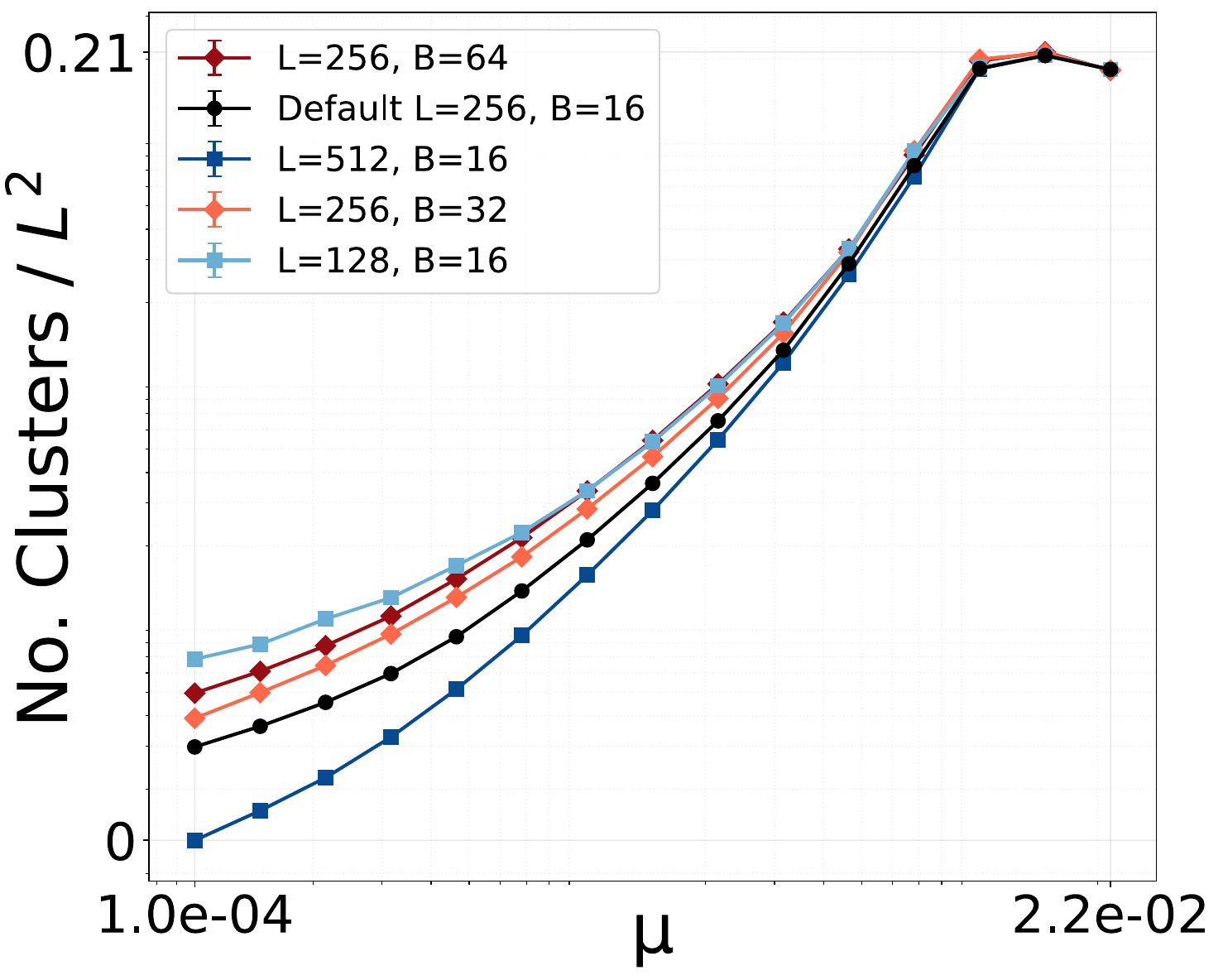}
    \includegraphics[width=0.32\linewidth]{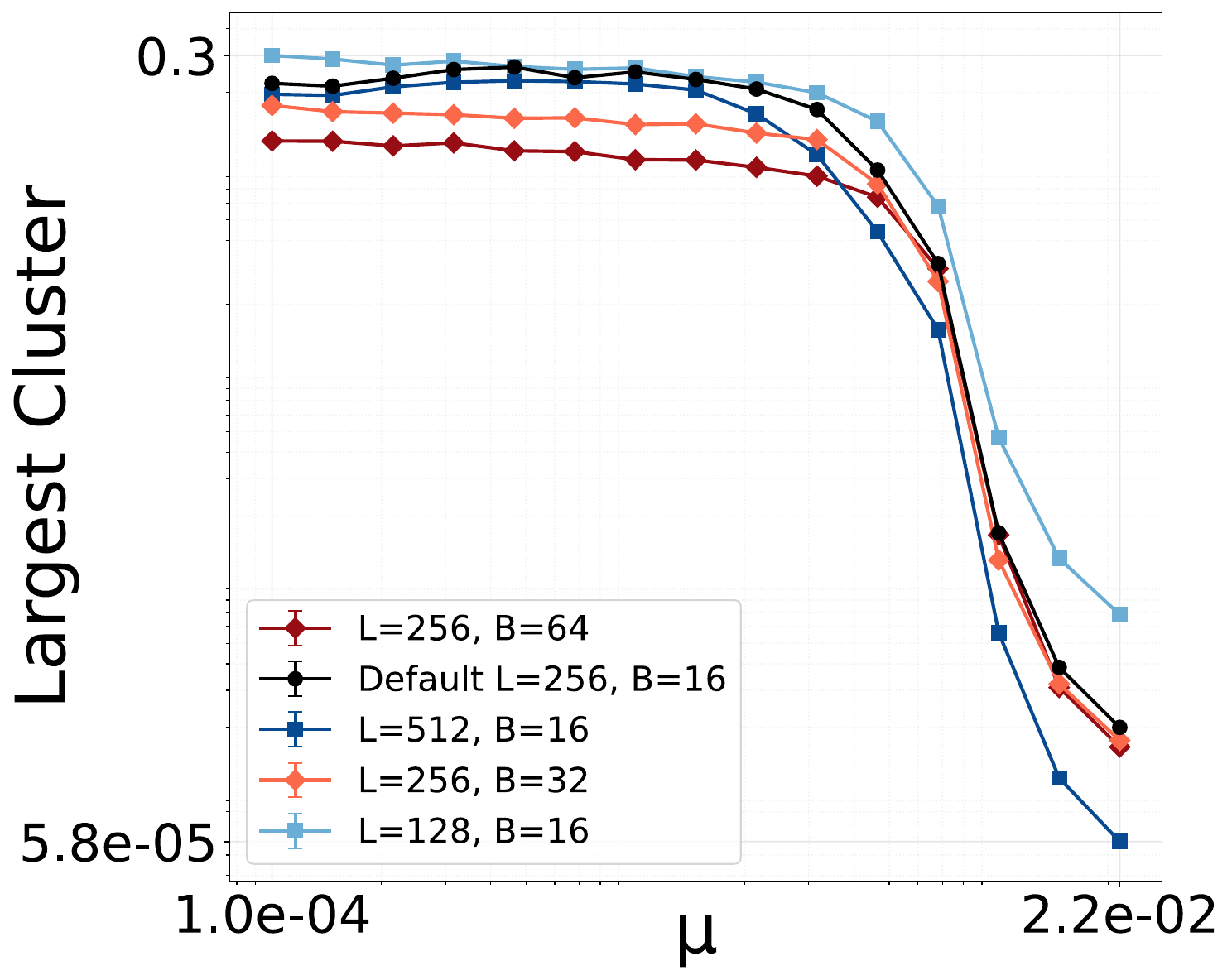}
    \caption{Plots of order parameters vs $\mu$, on a log-log plot/ Left: The number of clusters, Middle: The number of clusters normalized by $L^2$, Right: The largest cluster size, for $\gamma$=$\alpha$=1. Blue lines represent different system sizes $L$, red lines represent bitstring length $B$ (with $\mu$ scaled inversely appropriately). Bitstring length appears to not make a large difference to the number of clusters. Chasnging $L$ causes a large deviation when $\mu$ is large, due to the error catastrophe region, since every agent can be mutated. This can be better fit by rescaling by $L^2$, as in the middle.}
    \label{fig:mu-scaling-LB}
\end{figure}